\documentstyle[12pt]{article}
\begin{document}
\parindent 2em

\begin{titlepage}
\begin{center}
{\LARGE Superconducting transition induced by columnar disorder 
in strong magnetic field} 
\vspace{25mm}

Igor F. Herbut

Department of Physics and
Astronomy, 
University of British Columbia, 
Vancouver, BC V6T 1Z1, Canada

\end{center}
\vspace{10mm}

\noindent
{\bf Abstract:} The superconducting transition in presence of 
strong columnar disorder parallel to the magnetic field 
is considered. A solvable model appropriate for description of 
the broad crossover regime 
towards the true "glassy" critical behavior is constructed, and the 
behavior of the thermodynamical quantities and of the 
Edwards-Anderson order parameter
near transition is obtained. The critical exponents for the 
correlation lengths 
orthogonal and parallel to the magnetic field are the same and in 
agreement with the experimental values. The dynamical critical 
exponent is $z=2$, 
also in agreement with the measured value. Several perturbations 
to the solvable model are considered and shown to be irrelevant for the 
critical behavior. It is argued that there exists an optimal 
density of defects
at which the tansition temperature at given magnetic field reaches its 
maximum.

\end{titlepage}

\section{Introduction}
The problem of superconducting transition in the magnetic field 
in presence of strong 
disorder has been a subject of great interest in recent years. 
 From the point of view of applications, there is a need to know how to 
introduce defects into a superconducting material in a way that would 
maximize pinning of vortices and therefore 
increase critical currents \cite{1}. On the theoretical side, 
understanding of the 
phase transitions in presence of quenched disorder and 
description of the low-temperature 
disordered phases have always been  challenging problems, and  often 
required novel physical concepts and mathematical techniques.  
A prime example of this is the theory of spin glasses. For the 
superconducting transition in the magnetic field, a  
variety of novel low-temperature 
phases have been proposed, differing in the cases 
of point like \cite{2,3} and line-like disorder \cite{4,5,6}.
In the present paper we discuss a theory 
of superconducting transition at high magnetic fields 
 in presence of columnar (line-like) defects \cite{6}.  
We have in mind a three-dimensional, anisotropic, 
strongly type-II superconductor (YBCO, for example) in typical 
magnetic fields of $\sim 1T$, irradiated by a flux of some heavy ions 
with energies $\sim 1 GeV$ and with trajectories parallel 
to the external magnetic field. If the thickness of the sample in the 
direction of the beam is $\sim 10 \mu m$, the ions are able to 
penetrate through the entire material, leaving behind the continuous 
tracks of damaged superconductor of diameter $d\sim 50$\AA. 
In the absence of disorder, the high-field fluctuations 
of the order parameter, $\psi (\vec r)$, are
strongly enhanced by formation 
of Landau levels (LLs) for Cooper pairs. 
Such fluctuations lead to $D\to D-2$-dimensional reduction
in the pairing-susceptibility, $\chi_{sc}(\vec r,\vec r')$, 
and, strictly speaking, 
 eliminate the superconducting (Abrikosov) transition for $D=2,3$
\cite{dw}. The Abrikosov phase is then
replaced by a new fluctuation-induced state, the
density-wave of Cooper pairs (SCDW), in which the thermal average $\langle 
|\psi (\vec r)|^2\rangle$ has a weak modulation not necessarily 
accompanied by a long-range phase coherence \cite{dw}. 
The chief effect of disorder is to remove the LL degeneracy and so 
to restore a possibility of a true superconducting transition. 
In this sense the superconducting transition that we will 
consider is {\it induced} by the 
presence of disorder. $\chi_{sc}$ 
can now diverge at some finite temperature, $T_{sc}(H)$, determined
by the strength of disorder. The superconducting transition 
corresponds to the 
Bose condensation of Cooper pairs into the lowest energy eigenstate 
of the random potential, which, we will argue, extends over the whole 
sample in situations of experimental interest. 
Furthermore, for experimentally relevant parameters,
$T_{sc}(H)$ can be far above the SCDW transition line \cite{3dline}
over much of
the $H-T$ phase diagram, allowing us to treat the fluctuations
that produce SCDW in an approximate way.

In the following sections we first define a 
model for superconductor in a magnetic field parallel 
to the columnar defects, which represents 
 a realistic description of the 
problem for certain ranges of magnetic fields and densities of 
columnar defects. 
It is demonstrated that the model  is exactly solvable. 
The solution exhibits ``dimensional transmutation",
{\it i.e.} the effective dimensionality of the transition changes 
continuously as a function of magnetic field. This effect
is a direct consequence of analytic properties of
LL wavefunctions and is a signature of the high-field limit.
We determine the transition line in $H-T$ phase 
diagram, the Edwards-Anderson order 
parameter, and the behavior of 
specific heat and magnetic susceptibility in 
vicinity of the transition. There are similarities between the transition 
considered here and the one in the spherical model for spin-glasses 
\cite{spinglass}. Using the information on diffusion in strong 
magnetic field that comes from the studies of the quantum Hall effect, 
we determine the critical exponents for the correlation lengths and the 
dynamical critical exponent. The calculated exponents  
are in good agreement with the experimental values \cite{5}. 
We then consider perturbations to our exactly solvable model.  
In particular, we discuss the breakdown of the model for higher 
density of defects. It is argued that  disorder then becomes less 
efficient in removing LL degeneracy, enabling fluctuations again to suppress 
superconducting transition temperature, similarly to what happens in the 
homogeneous case. This naturally suggests the existence of the 
optimal dose of irradiation 
which produces the highest transition temperature at given magnetic field. 
We conclude by reviewing the salient points of our model 
and by briefly reanalyzing the approximations built into it. 

\section{Statement of the problem and the solvable model}

We are interested in strongly anisotropic layered superconductors 
described by the Ginzburg-Landau (GL) Lawrence-Doniach
model, with magnetic field perpendicular to the layers. 
Fluctuations in magnetic field are neglected (GL parameter 
$\kappa \gg 1$). We 
focus on the {\it high-field limit},
where the LL structure of Cooper pairs 
dominates the fluctuation spectrum: This is the case for fields above 
$H_{b} \approx (\theta/16)H_{c2}(0)(T/T_{c}(0))$, where $\theta$ is the 
Ginzburg fluctuation parameter \cite{7}. (For instance, in BSCCO 2:2:1:2, 
$\theta\approx 0.045$ and $H_{b}\approx 1$ Tesla.) In this regime, 
the essential features of the physics are captured by retaining only 
the lowest Landau level (LLL) modes. Generally, the 
partition function 
is $Z=\int D[\psi^{*}, \psi] \exp(-S)$, and 
\begin{eqnarray}
S=\frac{b}{T} \sum_{n=1} ^{N_{L}}\int d^{2}\vec{r} 
[ \eta |\psi_{n}(\vec{r})-
\psi_{n+1}
(\vec{r})|^{2} +  \nonumber \\
(\alpha '(T,H)+ 
\lambda \sum_{i} V(|\vec{r}-\vec{r_{i}}|)) 
|\psi_{n}(\vec{r})|^{2} + \frac{\beta}{2}
|\psi_{n}(\vec{r})|^{4}], 
\end{eqnarray}
 where $\alpha '(T,H) =a(T-T_{c2}(H))$, $b$
is the effective layer separation, $n$ is the layer index 
 and $a$, $\beta$ and 
$\eta$ are phenomenological parameters. 
The magnetic field is
assumed to be parallel to columnar defects, which are 
modeled by an effective potential $V(\vec{r}-\vec{r}_{i})>0$, 
peaked at $\vec{r}=\vec{r}_{i}$ with a width  
comparable to the diameter of the columns $d$. The parameter $\lambda>0$ 
represents the effective strength of disorder, and its positivness 
reflects the fact that the superconducting temperature is 
locally suppressed  by damaging the material. 
Random variables in the problem are two-dimensional coordinates of 
defects, $\{\vec{r_{i}}\}$. We assume that the positions of columns of 
damaged superconductor are uncorrelated, i. e. that they are 
distributed according to 
the Poisson distribution 
$P_{N}(\vec{r_{1}},...\vec{r_{N}})=(e^{-\rho A} \rho^{N})/N! $
where $P_{N}$ is the probability for finding $N$ impurities at the 
positions $\vec{r_{1}},...\vec{r_{N}}$, $A$ is the area of the system 
and $\rho$ is the concentration of impurities \cite{8}.

The partition function in Eq. 1 is quite general, 
and defines the problem we want to study. 
To proceed, note that in certain range of 
parameters one may make two simplifying assumptions. 
First, if the magnetic length $l$ 
and the average distance between defects are both much larger than 
the diameter of the columns $d$, we may assume 
that $V(\vec{r})=\delta(\vec{r})$. 
At fields of $\sim 1T$, $l\approx 200$\AA~,  and for moderate doses 
of $10^{9}-10^{11} ion/cm^{2}$ these  conditions are reasonably 
satisfied. This approximation makes the random potential 
which describes the disorder completely 
uncorrelated in space, and facilitates the exact treatment. 
Second, after 
rescaling the fields 
and the lengths as $(2b\beta 2\pi l^{2}/T)^{1/4} \psi \rightarrow \psi$, 
$r/(l\sqrt{2 \pi})\rightarrow r$, 
the quartic term can be rewritten as 
\begin {equation}
\frac{1}{4}\sum_{n}\int d^{2}\vec{r} |\psi_{n}|^{4} = 
\frac{1}{4N}\sum_{n} \beta_{A}(n) (\int d^{2}\vec{r} |\psi_{n}|^{2})^{2}, 
\end{equation}
where $\beta_{A}(n)=(N \int |\psi_{n}|
^4)/(\int |\psi_{n}|^{2})^{2}$ is the generalized Abrikosov ratio 
corresponding to configuration 
$\psi_{n}(\vec r)$, and $N=A/2\pi l^2$ is the degeneracy of the 
LLL. We now observe that $\beta_A (n)$
is only weakly dependent on the actual configuration, the well known 
example being the small difference in $\beta_{A}$ between 
triangular and square lattice of zeroes \cite{7}. 
Thus, we may approximate $\beta_{A}(n)$ in the quartic term by 
a constant of order unity. This replaces 
the local quartic term in the general theory (1)  
by an interaction still diagonal in layer indices, but 
infinitely ranged within a layer. This approximation neglects
the lateral fluctuations that produce the SCDW
transition \cite{dw}, and without disorder this theory has no 
finite temperature phase transition below four dimensions. Thus,  the 
disorder is assumed strong on the fine energy scale set by the 
local quartic term, but weak compared to the energy scale 
of LL separation, so 
that the LLL approximation is sensible. 
The neglect of weak lateral correlations 
is justified if $T_{sc}(H)$ is far above the
SCDW transition line \cite{3dline}. 
In that case the SCDW fluctuations start to matter  
only very close to the glassy transition and can be ignored in most
realistic situations. Since superconducting and SCDW transitions
arise from two distinct mechanisms, the respective transition lines
scale differently in the $H-T$ phase diagram and, for moderate
disorder, we are assured of a wide crossover region near $H_{c2}(T)$
where the neglect of SCDW fluctuations in our model is justified.

\section{Transition line and the order parameter}

After the $\beta_A (n)\to \langle\beta_{A}\rangle \sim 1$ substitution the
thermodynamics of the model becomes exactly 
solvable \cite{6}. We first introduce 
variables $\{x_{n}\}$ to decouple the quartic term 
and integrate over the fields $(\psi^{*},\psi)$. This leads to
$Z=\int \prod_{n} dx_{n}\exp(-NS')$, where
\begin{eqnarray}
S'= -\sum_{n} \frac{x_{n}^{2}}{\langle \beta_{A}\rangle} +
 \int_{0}^{\infty} dV \rho_{f}(V) 
Tr_{(n,m)} \nonumber \\
 \ln [g_{\eta}(2\delta_{n,m}-\delta_{n,m-1}-
\delta_{n,m+1})+
(g_{\alpha}+x_{n}+g_{\lambda}V)\delta_{n,n}]. 
\end{eqnarray}
We drop the terms coming from the rescaling of $\psi_n (\vec r)$ and
introduce dimensionless combinations of 
GL parameters $g_{\eta,\alpha,\lambda}=
\{\eta,\alpha ',\lambda/2\pi l^{2} \}\times\sqrt{
(b\pi l^{2})/(T\beta)}$.  The density of states for an uncorrelated 
random potential in the LLL can be found 
exactly by using the supersymmetric formalism \cite{9}. For
the Poisson infinitely short-range 
scatterers it is given by the integral: 
\begin{equation}
\rho_{f}(V)=\frac{1}{\pi}Im\frac{d}{dV}ln \int_{0}^{\infty} dt \exp(iVt-
f\int_{0}^{t}\frac{dy}{y}(1-e^{-iy})), 
\label{4}
\end{equation}
where $f=\rho 2\pi l^{2}=H_{\Phi}/H$, and $H_{\Phi}$ is the matching field, 
at which there is precisely one defect per unit of flux. 
 In the thermodynamic limit $N\rightarrow\infty$,
 the partition function 
is completely determined by the saddle-point of $S'$. 
Assuming that the saddle point is at $x$ independent of the 
layer index, we finally write the free energy above the critical 
temperature 
\begin{equation}
\frac{F}{NN_{L}T}=-\frac{x^{2}}{\langle \beta_{A}\rangle}+\frac{1}{2} \int_{-1}^{1} dk
\int_{0}^{\infty}dV\rho_{f}(V)\ln \left[g_{\eta}e(k)+
g_{\alpha}+x+g_{\lambda}V\right],
\end{equation}
where $e(k)=1-\cos (k)$ and $x$ is determined by the solution of 
\begin{equation}
x=\frac{\langle\beta_{A}\rangle}{4}\int_{-1}^{1} dk \int_{0}^{\infty} \frac{\rho_{f}(V) dV}
{g_{\eta}e(k)+g_{\alpha}+x+g_{\lambda}V}. 
\label{6}
\end{equation} 

In Eq. 6 it is important to know 
the behavior of density of 
states at low energies. For $f<1$, density of states has a delta-function 
singularity at $V=0$, while for $f>1$,  $\rho_{f}(V)\sim V^{f-2}$  
when $V\rightarrow 0$ \cite{9,10}. 
The transition line, $T_{sc}(H)$, in the
$H-T$ diagram is determined by  Eq. 6 and 
$x+g_{\alpha}=0$, which corresponds to condensation of Cooper pairs 
into $k=0$ and $V=0$ eigenstate of the random potential. 
Without disorder the LLL is completely degenerate and $\rho (V)=\delta(V)$.
The integral on the right hand side of Eq. 6 is then infrared 
divergent when $x+g_{\alpha}=0$, and there can be no finite temperature 
phase transition. With the columnar disorder present, from the 
behavior of the density of states it is easily seen that 
there will be a non-zero transition 
temperature only if concentration of impurities and magnetic field are 
such that $f>3/2$. Below this value of $f$ 
LLL degeneracy  is not sufficiently lifted by the random 
potential and thermal fluctuations 
prevent a finite temperature phase transition
in our model. $f=3/2$ 
determines the effective lower critical dimension for our model.  
After introducing dimensionless quantities $t=T/T_{c}(0)$, 
$h=H/H_{c2}(0)$ and $\lambda '=\lambda H_{c2}(0)/\phi_{0}aT_{0}$, 
where $\phi_{0}$ is the flux quantum, we perform the integration over wave-vector 
$k$ in Eq. 6 to obtain the expression for transition temperature: 
\begin{equation}
t_{sc}(h)=(1-h) \left[1+\frac{\theta\langle \beta_{A}\rangle}{2\lambda '} 
\int_{0}^{\infty}\frac{\rho_{f}(V) dV}{\sqrt{V^{2}+(2\eta V)/(h\lambda ' 
aT_{0})}}\right]^{-1}.
\end{equation}
Notice that when $\lambda '\rightarrow 0$ we have $t_{sc}(h)\rightarrow 0$, 
while for increasing $\lambda '$, transition temperature 
$t_{sc}(h)$ increases. 
Also, with increasing density of columnar defects 
(increasing parameter $f$), $t_{sc}(h)$ increases. This is 
related to the experimental observation \cite{1}
 that the irreversibility line shifts 
to higher temperatures with increasing dose of irradiation with heavy ions.

As temperature drops below $t_{sc}(h)$, $x$ remains 
at the value it had at the 
transition. There is now a macroscopic occupancy 
of the lowest energy state at $V=0$ and $k=0$. 
As is well known, condensation into this state is a meaningful concept 
only if 
the state is {\it extended}. Condensation into a 
localized lowest energy state would 
imply a diverging Abrikosov ratio $\langle\beta_{A}\rangle$ 
below the transition, making the transition impossible.
It is a 
special feature of this problem that the lowest lying state must indeed 
be extended for certain range 
of impurity concentrations. The density of 
states, Eq. 4, changes from being 
infinite at $V=0$ when $f<2$, to being zero 
when $f>2$. Thus, for fields and 
impurity concentrations such that parameter $f<2$, true extended states 
(which always exist in the LLL \cite{11}) must 
lay at the bottom of impurity band, since the number of states there 
diverges. The change of behavior in the density of 
states at $f=2$ could be caused by the fact that 
the mobility edge shifts to 
positive energies at some $f_0>2$, leaving spread-out but localized states 
at $V=0$, which now becomes the tail of the distribution.
Numerical diagonalization studies indicate 
that mobility edge is indeed located
near the band center for $f>4$ \cite{12}.  
Thus, strictly speaking, our model is appropriate for
$f<f_0$. However, even for $f$ above but close to $f_0$,
which is often the case for fields and concentrations 
of experimental interest, the 
states at $V=0$ are still near mobility edge and will appear 
extended in a finite size sample. On this basis, we expect that
useful information about the transition can still be obtained 
within our model. 

With these cautionary remarks in mind,  
the natural order parameter is the thermal average 
of the component of $\psi_n(\vec r)$ corresponding to the eigenvalue with 
$V=0$ and $k=0$. This is 
$\langle \psi_{0,0} \rangle =(NN_{L}(g_{\alpha}|_{t=t_{sc}(h)}-g_{\alpha})/
\langle\beta_{A}\rangle)^{1/2}$.
The disorder-averaged value of the field is 
$\overline{\langle\psi_{n}(\vec{r})\rangle}
=0$, due to random phases of the 
state $\phi_{V=0}(\vec{r})$.
Under the assumption that the lowest 
state is extended through the sample, 
$\overline{|\phi_{V=0}(\vec{r})|^{2}}\approx 1/N$; 
the Edwards-Anderson order parameter \cite{13}
 $q_{EA}=\overline{|\langle\psi_{n}(\vec{r})\rangle|^{2}}$ then equals 
\begin{equation}
q_{EA}=\frac{2}{\langle\beta_{A}\rangle}(g_{\alpha}|_{t=t_{sc}(h)}-g_{\alpha})
\end{equation}
below $t_{sc}(H)$, and is zero above. Thus, 
$q_{EA}\propto (t_{sc}(h)-t)^{2\beta}$, with the exponent $\beta=1/2$.
The free energy below  $t_{sc}(h)$ is 
\begin{equation}
\frac{F}{NN_{L}T}=-\frac{g_{\alpha}^{2}}{\langle\beta_{A}\rangle} + \frac{1}{2}
\int_{-1}^{1}dk\int_{0}^{\infty}\rho_{f}(V) dV \ln {(g_{\eta}e(k)
+g_{\lambda} V)}.
\end{equation} 

\section{Correlation lengths and dynamical exponent}

To calculate the exponents that determine the 
divergence of correlation lengths parallel and perpendicular
to the field  we first note that 
from Eq. 6 and the definition of critical line it follows 
\begin{eqnarray}
(g_{\alpha}+x)[1+\frac{\langle\beta_{A}\rangle}{4}
\int_{-1}^{1}dk\int_{0}^{\infty}
\frac{\rho_{f}(V)dV}{(g_{\eta}e(k)+g_{\lambda}V)(g_{\eta}
e(k)+g_{\lambda}V+g_{\alpha}+x)}] \nonumber \\
=g_{\alpha}-g_{\alpha}|_{t=t_{sc}(h)}.
\end{eqnarray}
The integral in the last equation diverges for $f<5/2$ 
as $(g_{\alpha}+x)^{f-5/2}$ 
when the transition line is approached from above, 
 and it is finite for $f>5/2$. Thus, we obtain 
 $(g_{\alpha}+x)\propto (t-t_{sc})^{1/(f-3/2)}$ for $f<5/2$ and 
$(g_{\alpha}+x)\propto (t-t_{sc})$ for $f \ge 5/2$. The same 
behavior follows if the transition line is approached along the line of 
constant temperature. This determines the 
value of the exponent for the correlation length parallel to the 
field $\xi_{\|}\propto (t-t_{sc}(h))^{-\nu_{\|}} $ : 
\begin{equation}
\nu_{\|}=1/(2f-3),
\end{equation}
 for $f<5/2$ and the 
classical value $\nu_{\|}=1/2$ for $f> 5/2$. 
The concentration corresponding to
$f=5/2$ determines the effective upper critical dimension in the problem. 
We now turn to correlation length perpendicular to the field, 
$\xi_{\bot}\propto[t-t_{sc}(h)]^{-\nu_{\bot}}$ and study the susceptibility 
associated with Edwards-Anderson order parameter, 
\begin{equation}
\chi_{EA}(\vec{r}-\vec{r'})\equiv\overline{\langle\psi_{n}^{*}(\vec{r})\psi_{n}
(\vec{r'})\rangle\langle\psi_{n}(\vec{r})\psi_{n}^{*}(\vec{r'})\rangle}.
\end{equation}
After expanding the field operators in the eigenbasis of 
random potential we obtain 
\begin{equation}
\chi_{EA}(\vec{r}-\vec{r'})=\int \frac{dV_{1}dV_{2}dk_{1}dk_{2} 
F(\vec{r}-\vec{r'},V_{1},V_{2})}
{(g_{\eta}e(k_{1})+g_{\lambda}V_{1}+g_{\alpha}+x)(g_{\eta}e(k_{2})+
g_{\lambda}V_{2}+g_{\alpha}+x)}
\label{12}
\end{equation}
where the function $F$ is the two-particle spectral density \cite{14}
\begin{equation}
F(\vec{r}-\vec{r'},V_{1},V_{2})=\overline{\sum_{i,j} \delta(V_{1}-V_{i}) 
\delta(V_{2}-V_{j}) \phi_{i}^{*}(\vec{r})\phi_{i}(\vec{r'}) 
\phi_{j}(\vec{r})\phi_{j}^{*}(\vec{r'})}
\end{equation}
and $\phi_{i}(\vec{r})$ are the eigenstates of the random potential. 
If we now introduce $V=(V_{1}+V_{2})/2$ and $\omega=(V_{1}-V_{2})/2$, for 
$V$ close to mobility edge and small $(q,\omega)$, the
Fourier transform of $F$ 
has a diffusive form \cite{14,15}
\begin{equation}
F(\vec{q},V,\omega)=\frac{\rho_{f}(V) q^{2} D(q^2/\omega)}
{\pi(\omega^{2}+q^{4}D^{2}(q^{2}/\omega))},
\end{equation}
where $D(q^{2}/\omega)$ is the  generalized ``diffusion constant".
Assuming this form for $F(\vec{q},V,\omega)$ and rescaling 
everything by the appropriate power of temperature in 
Eq. 12, we find:
\begin{equation} 
\nu_{\bot}=\nu_{\|}. 
\end{equation}
Surprisingly, the scaling turns out to be isotropic, in spite of 
the strong anisotropy in the  model. 
 
 The expression for two-particle spectral density yields 
immediately another important result. Since the energy and the 
momentum variables appear always in combination 
$\omega/q^2$, the dynamical exponent 
has the value:
\begin{equation}
 z=2.
\end{equation}
This may also be found by directly calculating 
the dc conductivity along the field using the 
time-dependent version of the model, which yields 
$\sigma_{zz}\propto \xi_{\|}$,     
the same as Aslamazov-Larkin result in three dimensions. 

The dependence of the correlation length exponents on magnetic field 
through parameter $f$ is the consequence of power-law behavior of the 
density of states close to the bottom of the band. A closer look at the 
the density of states $\rho_{f}(V)$ reveals it to be 
roughly constant except in a narrow 
region, typically less than 1\% of total bandwidth,
around $V=0$, where it either diverges of vanishes \cite{10}.
 One might 
expect that taking the effective potential $V(\vec{r})$ to have a finite 
range would tend to wash away this fine feature of the density of states. 
This suggests that the experimentally relevant situation corresponds 
to $f=2$ in our model, when the density of states is flat down 
to the lowest energies. Based on this observation one would expect that 
the observable values of the exponents are 
$\nu_{\bot}=\nu_{||}=1$. This expectation is met by the 
experimental results of Ref. 5: $\nu_{\bot}=1.0 \pm 0.1$, $\nu_{\|}=1.1
\pm 0.1$. The value of dynamical exponent is also in agreement with 
the measured  $z=2.2 \pm 0.2$ \cite{5}. 

\section{Magnetization and specific heat}
 
Magnetization per unit volume is found by differentiating the averaged 
free energy with respect to the magnetic field: 
\begin{equation}
\frac{M}{AN_{L}d}=-\frac{2T_{0}\sqrt{th}}{d\phi_{0}\sqrt{\theta}}\left(q_{EA}+
\frac{2x}{\langle\beta_{A}\rangle}\right).
\end{equation}
Below the transition line this coincides with the usual mean-field result. 
Above the transition line $q_{EA}=0$ and from Eq. 10 
it follows that at constant temperature close to the transition 
$(g_{\alpha}+x) \propto[h-h_{sc}(t)]^{1/(f-3/2)}$ when 
$f<5/2$ and $(g_{\alpha}+x) \propto [h-h_{sc}(t)]$ otherwise. Thus the magnetic 
susceptibility is a smooth function of the field at the transition 
for $f<2$, but has an upward  cusp 
for $2<f<5/2$ and a discontinuity for $f>5/2$. The size of this discontinuity 
depends on the location of the transition in the $H-T$ diagram. 
Differentiating the free energy twice with respect to temperature 
one obtains the specific heat. It is straightforward to show that 
at the transition it behaves the same way as susceptibility; smooth for 
$f<2$, has a cusp for $2<f<5/2$ and has the usual discontinuity for 
$f>5/2$. More precisely, both magnetic susceptibility and specific 
heat behave as $[t-t_{sc}(h)]^{-\alpha}$
for $3/2<f<5/2$, where $\alpha=(f-5/2)/(f-3/2)$.
The behavior of the specific heat, order parameter and 
correlation length in our model is related to the 
one obtained from O(2$N$) vector model in the limit 
$N\rightarrow \infty$ and in the {\it effective} dimension $D_{eff}=2f-1$.  

\section{Perturbations}

Having a solvable version of the general theory (1), we 
are in position to study small perturbations around it and 
check their relevance for the critical behavior. First, let us 
assume that the strength of disorder $\lambda$ is also allowed to 
fluctuate from one columnar defect  to another, i. e., the random  
potential in Eq. 1 is taken to be
\begin{equation}
\sum_{i} \lambda_{i} \delta(\vec{r}-\vec{r}_{i}), 
\end{equation}
with the random variable $\lambda_{i}$ distributed according to the 
distribution:
\begin{equation}
P(\lambda_{i})=\frac{1}{2 w}, 
\end{equation}
for $|\lambda_{i}-\lambda|<w$, $0<w<\lambda$, and zero otherwise. 
This random potential, being still spatially uncorrelated, allows 
an exact calculation of the density of states \cite{9}. 
One can show however \cite{10}, that the power-law behavior 
of the density of states at low energies 
is not changed by allowing finite width $w$, 
which implies the same critical exponents as already obtained. 
This perturbation is therefore irrelevant.

 A more complicated extension of the model 
is to assume a particular form of  
three-dimensional 
disorder represented by a random potential in Eq. 1: 
\begin{equation}
\lambda_{n}\sum_{i} \delta(\vec{r}-\vec{r}_{i}), 
\end{equation}
where the effective strength of disorder $\lambda_{n}$ is the same for all 
defects within the $n$-th layer, 
but fluctuates from one layer to another according to 
a Gaussian distribution
\begin{equation}
P[\{\lambda_{n}\}]\propto \exp{-\sum_{n} (\lambda_{n}-\lambda)^2 / w^2}, 
\end{equation}
where the width of distribution $w$ is a small parameter, $w<<\lambda$. 
Since the model is not fully three, but rather $2+1$ 
dimensional, one can still study the phase 
transition by the method described here 
in combination with the replica formalism 
\cite{13}. Without going into details, the result is that the 
critical behavior is not changed by a small $w$, and the only effect 
of additional disorder along the field is a 
decrease in transition temperarature. 
We conclude that this perturbation is also irrelevant for the 
critical behavior. 

As emphasized earlier, the replacement of 
disorder potential representing disorder by  a sum of delta-functions 
is sensible only if the other two length scales in the problem, 
magnetic length and the average separation between the defects, are 
much larger than the average diameter of the columns. The latter 
condition obviously breaks down for higher doses of irradiation, 
when there is a considerable probability of overlap between the 
columnar defects. In fact, 
at higher densities of defects the trend of 
increase of the transition temperature with defect concentration 
should be reversed. 
To see this, consider first a small density of defects. Then 
our model is applicable, and as we already 
noted, at fixed magnetic field the 
transition temperature given by eq. 7 increases 
with increasing parameter $f$. This is because larger $f$ shifts the 
density of states towards higher energies, decreasing the integral 
in denominator in the Eq. 7. For high defect densities the model ceases to 
be useful, but it is still possible to qualitatively understand 
what happens by considering the limit of very large concentrations. 
 For very large density of defects, the effect of randomly distributed 
finite width scatterers is similar to having a potential which 
is almost constant in space. In such a potential the LLL is again nearly 
degenerate, only the average energy is shifted by some amount from 
where it was without any external potential. But this being the 
case, thermal fluctuations are again able to suppress the transition 
temperature because of the dimensional reduction! Thus,  for large densities 
of defects transition temperature must be decreasing with the 
density. This suggests that there is a field dependent optimal defect 
concentration at which the transition temperature reaches its maximum
value, still below the mean-field $T_{c2}(H)$. 
This would explain the observation of saturation of transition temperature 
with increase of dose of irradiation in ref. 16. 

\section{Summary and conclusions}
 
We have studied the high-field superconducting glassy transition 
induced by columnar disorder. It was argued that in certain 
range of magnetic fields and for not too large concentrations of defects 
the random potential that represents 
disorder may be assumed to be uncorrelated in space. 
By assuming 
the self-interaction of the fluctuating superconducting 
order-parameter to be 
infinitely ranged in directions perpendicular to the field we 
defined a model which describes the crossover regime towards 
the true glassy critical behavior, and which can be solved exactly. 
Within this model we have calculated the transition line, 
Edwards-Anderson order parameter, magnetization, specific heat, 
critical exponents for the correlation lengths and the dynamical exponent. 
The critical behavior of the model is shown to be isotropic, and the 
values of the critical exponents are in agreement with existing 
measurements. Several 
extensions of our model are considered, and some are shown to 
be irrelevant for the critical behavior. We presented a qualitative 
arguments for the behavior of transition temperature with 
density of defects, and argued for the existence of optimum 
concentration of defects at which the transition temperature reaches 
its maximum. 

The crucial simplifications of the general GL theory with columnar 
disorder in Eq. 1 which facilitated the exact solution of the model 
were the assumption of delta-function scatterers for 
representation of disorder and the neglect of weak SCDW correlations, 
i. e., $|\phi|^{4}$ interaction term was replaced by 
interaction infinitely long-ranged 
within layers. The first approximation is not a drastic 
one for small densities of defects; allowing for some range of 
random scatterers is likely to wash away the power-law behavior 
of density of states at low energies and make the critical behavior 
analogous to the $f=2$ case for our model. This picture is 
in agreement with the measured values of critical indices for 
correlation lengths \cite{5}. Neglect of SCDW correlations represents 
a more serious approximation, which necessarily breaks down 
sufficiently close to the critical temperature. The model solved here 
is thus best understood as a description of the crossover region before the 
true critical (glassy) behavior is reached. It is a disordered 
equivalent of spherical model for spin systems. Strong disorder 
however, guarantees a rather wide crossover region where our model 
is useful. This is supported by experimental observation of nearly 
isotropic scaling \cite{5}, while the general model in Eq. 1 would be 
most likely to exhibit an anisotropic critical exponents. 
Also, for calculations of non-universal quantities, such as the 
transition line in $H-T$ phase diagram, the present model augmented 
with more realistic potential $V(\vec{r})$ should suffice. It is 
therefore a useful step in quantitative understanding of superconducting 
transition in presence of correlated disorder in magnetic field. 

\section{Acknowledgments}

Most of the work described here was done in collaboration with 
Professor Zlatko Te\v sanovi\' c while the author was at Johns Hopkins 
University. The work there was supported by 
NSF Grant No. DMR-9415549. Work at University of British Columbia 
has been supported by Natural Sciences and Engineering Research Council 
of Canada and Izaak Walton Killam postdoctoral fellowships.

\end{document}